\documentclass[journal]{IEEEtran}
\usepackage[utf8]{inputenc}

\ifCLASSINFOpdf
  \usepackage[pdftex]{graphicx}
  \graphicspath{{../} {../pdf/} {../jpeg/}}
  \DeclareGraphicsExtensions{.pdf,.jpeg,.png}
\else
  \usepackage[dvips]{graphicx}
  \graphicspath{{../} {../eps/}}
  \DeclareGraphicsExtensions{.eps}
\fi

\usepackage{hyperref} 
\usepackage{amssymb}
\usepackage{amsmath}
\usepackage{amsthm}
\usepackage{bm}
\usepackage{float}
\usepackage{color}
\usepackage{colortbl}

\usepackage{threeparttable}
  
\title{A Data-Driven Uncertainty Quantification Method for Stochastic Economic Dispatch}
\author{
{Xiaoting Wang, Rong-Peng Liu, Xiaozhe Wang, Yunhe Hou, Fran\c{c}ois Bouffard}
\thanks{ This work was supported partially by Natural Sciences and Engineering
Research Council (NSERC) Discovery Grant, NSERC RGPIN-2016-04570 and partially by the Fonds de Recherche du Qu\'{e}bec-Nature et technologies under Grant FRQ-NT PR-253686. }
}
\date{November 2020}

\begin{document}

\maketitle

\begin{abstract}
This letter proposes a data-driven sparse polynomial chaos expansion-based surrogate model for the stochastic economic dispatch problem considering uncertainty from wind power. The proposed method 
can provide accurate estimations for the statistical information (e.g., mean, variance, probability density function, and cumulative distribution function\color{black}) for the stochastic economic dispatch solution efficiently without requiring the probability distributions of random inputs. Simulation studies on an integrated electricity and gas system (IEEE 118-bus system integrated with a 20-node gas system\color{black})  
are presented, 
demonstrating the efficiency and accuracy of the proposed method compared to the Monte Carlo simulations. 

\end{abstract}

%
\vspace{-0.1in}
\section{Introduction}
%
The intermittency of power generation from renewable energy sources (RES) results in great challenges in power system daily operation like economic dispatch (ED).  
To address the challenge of uncertainty in the ED problem, various formulations have been proposed such as stochastic programming, i.e., stochastic ED (SED), 
robust optimization formulation, 
chance-constrained optimization, 
etc. A large number of Monte Carlo (MC)-based simulations are needed to solve the optimization problem using realizations from stochastic  models, which unfortunately yields prohibitive computational efforts \cite{Safta2017}.  

To lower the computation burden, uncertainty quantification methods have been applied to the SED problem. Safta et al. \cite{Safta2017} applied the polynomial chaos expansion (PCE) method to build a surrogate model for the SED problem, which can  achieve accurate results efficiently with fewer samples compared to MC-based methods. Li et al. \cite{Li2019} further adopted a sparse PCE (SPCE) method 
to alleviate “the curse of dimensionality”.  
However, the surrogate models in \cite{Safta2017}, \cite{Li2019} are constructed by assuming that the random inputs follow Gaussian distributions, which may not be true in practice. More recently, a Gaussian process emulator-based \color{black}
approach was proposed in \cite{Hu2021} to solve the SED problem. Yet, it has been discussed in \cite{Rajabi2019} that the PCE method may outperform the Gaussian process emulator-based method when the probability distribution of the response (e.g., the SED solution) tends to be multimodal.  

In this letter, we propose a data-driven sparse PCE (DDSPCE) method 
that builds a surrogate model directly from a raw data set of random variables,  without any prior assumptions on the marginal distributions of the random variables and output responses. The DDSPCE-based surrogate \color{black}model can accurately estimate the statistical information (e.g., mean, variance, probability density function (PDF), cumulative distribution function (CDF)) of the SED solution efficiently (e.g., $33$ times faster than MC simulations \color{black} on the  integrated IEEE 118-bus power system and 20-node gas system \cite{Liu2021}), 
even when the PDF of the SED solution is multimodal. 

\vspace{-8pt}
\section{Formulation of Stochastic Economic Dispatch}\label{sec:SED_formulation}
%

The two-stage models are typically used to solve the stochastic unit commitment (UC) problem, where the first stage is to find the UC decisions in the day-ahead electric market, and the second stage is to find the dispatch decisions \cite{Liu2021}. \color{black} 
In this letter, we assume that  the first stage UC \color{black} decisions for conventional generating units have already been determined from the day-ahead UC model  similar to \cite{Safta2017,Hu2021}. \color{black} 
Our goal is to solve the multi-period ED problem with fixed UC decisions while considering the uncertainties from RES. 
The uncertainties introduced by the power output of renewable generating units turn the  production cost $Q(\bm{P_g},\bm{P}_{\bm{w}})$ into a random variable, where $\bm{P_g}$ denotes 
the generator power output vector; $\bm{P}_{\bm{w}}$ is a vector of random variables, e.g., wind power generation in this paper. The multi-period 
SED problem under fixed UC decisions 
can be represented by: 
%
\small{
\begin{equation}
\setlength{\abovedisplayskip}{1pt}
\setlength{\belowdisplayskip}{1pt}
\label{eq:SED_Obj}
   Q(\bm{P_g},\bm{P}_{\bm{w}})= \min_{\bm{P_g}} \sum_{t\in T}\sum_{g\in G} C_{g}(P_{g}^{t})
\end{equation}   
\vspace{-0.23in}
\begin{subequations}
\setlength{\belowdisplayskip}{0pt}
\label{eq:constraints}
\begin{flalign}
& \mathrm{s.t.}  \notag \\
 & \sum_{g \in G}P_{g}^{t} + \sum_{w \in W} P_{w}^{t} = \sum_{d \in D} P_{d}^{t}\quad \forall t\in T \label{eq:Power_bal}\\
& \underline{P_{l}} \leq \sum_{g \in G} k_{lg} P_{g}^{t}+\sum_{w\in {W}} k_{lw} P_{w}^{t} -\sum_{d \in D} k_{ld}P_{d}^{t}\leq \overline{P_{l}} \quad \forall t\in T \label{eq:DC_flow} \color{black}\\
& P_{g}^{\mathrm{min}}x_{g}^{t}\leq P_{g}^{t}\leq P_{g}^{\mathrm{max}} x_{g}^{t} \quad  \forall g\in G, t\in T \label{eq:Gen_Limit}\\
& -R_{g}^{RD}x_{g}^{t}-R_{g}^{SD}(x_{g}^{t-1}-x_{g}^{t})-P_{g}^{\mathrm{max}}(1-x_{g}^{t-1}) \notag\\ 
& \leq P_{g}^{t}-P_{g}^{t-1} 
\leq R_{g}^{RU}x_{g}^{t-1} + R_{g}^{SU}(x_{g}^{t}-x_{g}^{t-1}) \notag\\  
& + P_{g}^{\mathrm{max}}(1-x_{g}^{t}) \quad \forall g\in G, t\in T \label{eq:Ramp}
\end{flalign}
\end{subequations}
}
\normalsize
where (\ref{eq:SED_Obj}) is the objective function, i.e., to minimize the total production cost; 
$t$ is the specific time period in the time periods set $T=\{1,\cdots,T_m\}$ (e.g., 24-hour period in the simulation study); $g$ is the generator index and $G$ is the generator set. 
The operational and physical constraints are given in (\ref{eq:constraints})  based on direct current (DC) power flow model. Equation (\ref{eq:Power_bal}) is the power balance constraint, where 
$P_d^{t}$ is the $d$-th load demand at time $t$.
Constraint \color{black} (\ref{eq:DC_flow}) 
denotes \color{black}the power flow limits, where $k_{lg}$, $k_{lw}$ and $k_{ld}$ are the sensitivity coefficients for the $l$-th transmission line with respect to the traditional generator $g$, wind generator $w$ and load $d$, respectively \cite{Liu2021}. 
$\underline{P_{l}}$ and $\overline{P_{l}}$ are thermal limits of transmission line $l$. \color{black}  
Constraint \color{black} (\ref{eq:Gen_Limit}) denotes the generation capacity limits with 
$x_{g}^{t}$ being the pre-determined UC decision for generator $g$ at time $t$. Constraint  (\ref{eq:Ramp}) describes the  ramping capability constraint of generator $g$, where $R_{g}^{RD}$ and $R_{g}^{RU}$ denote the ramping down and up rate; $R_{g}^{SD}$ and $R_{g}^{SU}$ denote the shut-down and start-up ramp rate. Note that in this letter the gas system is integrated into the power system for case study. Additional constraints of gas network  are considered (see Appendix), 
yet the essence of the SED problem is not affected. Please check \cite{Liu2021} for more details. \color{black}


\vspace{-12pt}
As can be seen, the SED problem by nature is a complex constrained optimization problem. 
MC simulations are typically adopted to estimate the cost from a finite set of  realizations from stochastic models of load variation, wind generation, etc., which nevertheless lead to high computation efforts even with scenario reduction techniques. 
In this letter, we propose a DDSPCE-based surrogate model to estimate the expected minimum cost using much less samples compared to MC-based approaches. 
The proposed method can effectively estimate the statistics 
as well as the PDF and CDF of the minimum cost $Q(\bm{P_g},\bm{P_w})$ purely from the wind power data. 

\vspace{-0.13in}
\section{Data-Driven based Polynomial Chaos Expansion}
\vspace{-3pt}


In this section, a DDSPCE-based surrogate model, i.e., a linear combination of multivariate orthogonal polynomials, will be developed to represent the relationship between the input random variables $\bm{P_w}$ and the minimum cost $Q(\bm{P_g},\bm{P_w})$. 
\vspace{-0.35in}
\subsection{Polynomial Chaos Expansion}

Consider an independent random vector $\bm{\xi}=\{\xi_1,...,\xi_M\}\in\mathbb{R}^{M}$, serving as the input to the system, then the model response $Y =f(\bm{\xi})$, being a function of $\bm{\xi}$, is also stochastic. In the SED problem, $\bm{\xi}$ can be obtained by decorrelating the wind power $\bm{P_w}$ through, for example, the principal component analysis (PCA), while the model response $Y$ corresponds to the minimum production cost $Q$. 
It is claimed by the Cameron–Martin theorem that the model output $Y$ can be expressed by $K$ numbers of the expansion terms constructed by the orthogonal polynomial basis of $\bm{\xi}$ \cite{Li2019}:
\small
\begin{equation}
\setlength{\abovedisplayskip}{0pt}
\setlength{\belowdisplayskip}{0pt}
\label{eq:PCEinfty}
    \begin{aligned}
    Y=f(\bm{\xi})\approx  \sum_{i =1}^{K}c_i \Phi_{i}(\bm{\xi})
    \end{aligned}
\end{equation}
\normalsize
where $c_i$ are the unknown expansion coefficients to be determined; $\Phi_{i}(\bm{\xi})$ are the multidimensional polynomial basis, which are orthogonal to the joint marginal distribution of $\bm{\xi}$. Note that when $K \rightarrow \infty$, the series in (\ref{eq:PCEinfty}) converges in the sense of the $\mathcal{L}^2$-norm. For practical implementation, $\Phi_{i}(\bm{\xi})$ is generally truncated to a finite number of expansion terms: 
\small
\begin{equation}
\setlength{\abovedisplayskip}{0pt}
\setlength{\belowdisplayskip}{0pt}
\label{eq:tensor_trunc} 
     \Phi_{i}(\bm{\xi})   = \prod_{j=1}^{M} \phi^{(i_j)}_{j}({\xi_j}) 
\end{equation}
\normalsize
%
where $\Phi_{i}(\bm{\xi})$ is formed by a tensor product of univariate orthogonal polynomial basis $\phi^{(i_j)}_{j}({\xi_j})$, $j=\{1,\cdots,M\}$. $\phi^{(i_j)}_{j}({\xi_j})$ is the one-dimensional polynomial basis of $\xi_j$ with degree $i_j$ at $i$-th expansion term. $i_j$ denotes the degree of $j$-th univariate polynomial basis at $i$-th expansion term. 

Note that 
for random inputs with thick tailed distributions or nonlinear correlations, the PCA may provide bias results. To overcome this, vine copula and Rosenblatt transform may be applied to model the dependence and decouple the components of the input data 
\cite{TorreE2019}. The kernal PCA can also be applied to handle nonlinear correlations. \color{black}

\begin{figure}[]
\setlength{\abovecaptionskip}{-0.12cm}
\setlength{\belowcaptionskip}{-0.9cm}
\centering
\includegraphics[width=0.43\textwidth]{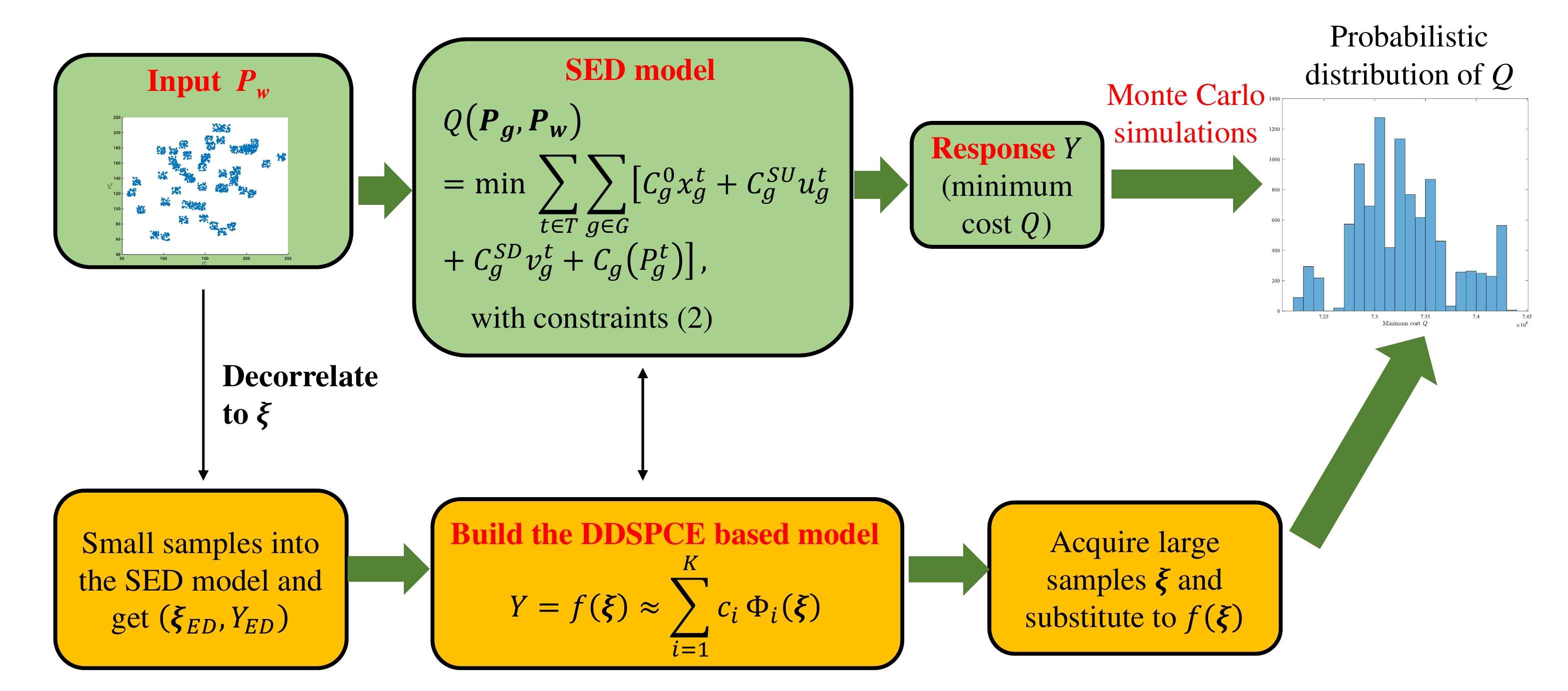}
\caption{Relation between the SED and DDSPCE-based surrogate model}
\label{fig:relation}
\vspace{-0.7cm}
\end{figure}

The essence of the proposed DDSPCE method and its relation to the original SED problem (\ref{eq:SED_Obj})-(\ref{eq:constraints}) is presented in Fig. \ref{fig:relation}. Instead of running MC simulations on the SED problem (\ref{eq:SED_Obj})-(\ref{eq:constraints}) using a large number of scenarios, we first evaluate $Q(\bm{P_g},\bm{P_w})$ using a small number of scenarios of $\bm{P_w}$. Next, we use the obtained inputs $\bm{P_w}$ and outputs $Q(\bm{P_g},\bm{P_w})$ to build a DDSPCE-based surrogate model (\ref{eq:PCEinfty}), a pure algebraic equation. Finally, we acquire a large number of scenarios of $\bm{P_w}$, the production cost  $Q(\bm{P_g},\bm{P_w})$ of which can be quickly calculated by substituting the detailed numbers into the established DDSPCE-based surrogate \color{black}model (\ref{eq:PCEinfty}). Compared to the SED problem  (\ref{eq:SED_Obj})-(\ref{eq:constraints}), the DDSPCE-based surrogate model is much faster to evaluate and thus can save significant computational efforts. 
To build the DDSPCE-based surrogate model, our main tasks are to  build the multidimensional polynomial basis $\Phi_{i}(\bm{\xi})$ using the available data of $\bm{P_w}$ and then to calculate the coefficients $c_i$ in (\ref{eq:PCEinfty}).

%
%

%
%

\vspace{-13pt}
\subsection{Constructing the Polynomials from Moments}
\vspace{-2.5pt}
For simplicity, we use $\phi^{(l)}_{j}({\xi_j})$ to replace $\phi^{(i_j)}_{j}({\xi_j})$ hereafter. The polynomial $\phi^{(l)}_{j}({\xi_j})$ of degree $l$ is defined as:
\small
\begin{equation}
\label{eq:univatiate_poly}
 \setlength{\abovedisplayskip}{0pt}
\setlength{\belowdisplayskip}{0pt}
    \phi^{(l)}_{j}({\xi_j}) = \sum_{k=0}^{l}p_{k,j}^{(l)}\xi_{j}^{k}, \quad l=\{0,1,\cdots,D\}
\end{equation}
\normalsize
where $p_{k,j}^{(l)}$ are the coefficients of polynomial $\phi^{(l)}_{j}({\xi_j})$ in the $k$-th degree.  $D$ is the order of the PCE-based model. \color{black}

Traditionally, the orthogonal polynomial $\phi^{(l)}_{j}({\xi_j})$ can be determined by the Weiner-Askey scheme based on some typical distribution types of $\xi_j$ \cite{Sheng2018}. However, exact knowledge of probability distribution of random inputs may hard to be acquired 
in practice, while raw data sets (e.g., wind power \cite{NREL2020}) are more likely to be available. In this letter, the main idea of the data-driven SPCE 
method is to calculate the coefficients $p_{k,j}^{(l)}$ based on a few statistical moments that can be directly estimated from limited data. The exact probability distributions of $\xi_j$ are not required to be known or even exist. $\xi_j$ can be either continuous, discrete, or mixed and can be specified either by raw data sets, histogram or probability distributions. 

The polynomial basis $\phi^{(l)}_{j}({\xi_j})$ needs to satisfy the orthogonality condition
\begin{small}
\begin{equation}
\setlength{\abovedisplayskip}{1pt}
\setlength{\belowdisplayskip}{2pt}
  \label{eq:Poly_basis_Orthogonal}  
  \int_{\Omega} \phi_{j}^{(m)}({\xi_j})\phi_{j}^{(l)}({\xi_j})d\Gamma({\xi_j}) = 0 \qquad \forall m \neq l
\end{equation}
\end{small}
where $\Gamma$ is the marginal cumulative distribution function of $\xi_j$. If we let the coefficients of leading terms for all polynomials to be $1$, i.e., $p_{l,j}^{(l)} = 1, \forall l$ in (\ref{eq:univatiate_poly})\color{black}, and utilize the definition of the $k$-th raw moment of $\xi_j$:
\begin{small}
\begin{equation}
\setlength{\abovedisplayskip}{0pt}
\setlength{\belowdisplayskip}{0pt}
\label{eq:moment_l}
    \mu_{k,j} = \int_{\xi_j \in \Omega} \xi_{j}^{k}d\Gamma(\xi_j)
\end{equation}
\end{small}
It can be shown that 
the coefficients $p_{k,j}^{(l)}$ in (\ref{eq:univatiate_poly}) can be calculated based on the following matrix equation \cite{Wang2021}:
\begin{small}
\begin{equation}
\setlength{\abovedisplayskip}{1pt}
\setlength{\belowdisplayskip}{1pt}
\label{eq:matrix_coefficients}
	\left[\begin{array}{cccc}
	\mu_{0,j} & \mu_{1,j} &\ldots & \mu_{l,j} \\
	\mu_{1,j} & \mu_{2,j} &\ldots & \mu_{l+1,j} \\
	\vdots & \vdots & \vdots & \vdots \\
	\mu_{l-1,j} & \mu_{l,j} & \ldots & \mu_{2l-1,j} \\
	0 & 0 & \ldots & 1
	\end{array}\right] \left[\begin{array}{c}
	p_{0,j}^{(l)} \\ p_{1,j}^{(l)} \\\vdots \\ p_{l-1,j}^{(l)} \\ p_{l,j}^{(l)}
	\end{array}\right] = \left[\begin{array}{c}
	0\\0\\ \vdots \\ 0\\1
\end{array}\right]
\end{equation}
\end{small}
where $\mu_{m,j}$ is the raw statistical moments of $\xi_j$ with $m=\{0,\cdots,2l-1\}$, $j=\{1,\cdots,M\}$,  $l=\{0,1,\cdots,D\}$. \color{black} $\mu_{m,j}$ can be easily calculated from the given raw data set or probabilistic distribution of $\xi_j$. 
Besides, 
the raw moment matrix in (\ref{eq:matrix_coefficients}) has to be nonsingular. This condition is satisfied when finite statistical moments up to  $2D-1$ \color{black} order exist as well as $D$ \color{black} or more distinct values are included in the data set, if $\mu_{m,j}$ is evaluated directly from data.  Generally, $\bm{\xi}$ represented  by  a  data  set  can satisfy these  conditions easily, since the data points are finite and the degree $D$ is small (typically $\leq 5$). \color{black}  

Once the coefficients $p_{k,j}^{(l)}$ are solved from (\ref{eq:matrix_coefficients}), the polynomial basis $\phi^{(l)}_{j}({\xi_j})$ and the multidimensional polynomial basis $\Phi_{i}(\bm{\xi})$  can be obtained from (\ref{eq:univatiate_poly}) and (\ref{eq:tensor_trunc}) in sequence. To build the PCE model (\ref{eq:PCEinfty}), we need to solve the coefficients $c_i$. To this end, the orthogonal matching pursuit \color{black}(OMP) is applied to find the best polynomial sets and the corresponding coefficients.
\vspace{-0.19in}
\subsection{Calculating $c_i$ by Orthogonal Matching Pursuit (OMP)}
\vspace{-0.01in}
OMP is an iterative algorithm, which selects regressors that are most correlated to current approximation residual and adds them to the active set of basis in each iteration and then updates the coefficients $c_i$ for all active regressors by the ordinary least squares (OLS) method. During the selection, the leave-one-out (LOO) error estimator ((1.26) in \cite{Marelli2018}) is chosen as a criterion for the model order $D$ and the sparse candidate basis. \color{black} 
Detailed description of the OMP procedure can be found in \cite{Marelli2018}.  
The OMP algorithm can achieve a sparse PCE-based surrogate \color{black}model, which reduces the  effort while guaranteeing the accuracy.

\vspace{-0.22in}
\section{Simulation Studies}
\vspace{-0.12cm}
In this section, we test the proposed DDSPCE method on an integrated electricity and gas system (IEGS), i.e., the IEEE 118-bus system integrated with a 20-node \color{black} gas system. 
5 wind farms are added into the system at bus \{2, 33, 51, 81, 108\} using the NREL's Western Wind Data Set \cite{NREL2020}. The time period considered is 24 hours, e.g., $T=\{1,\cdots,24\}$. 
Thus, the wind generator output $\bm{P_w}$ is a $120$-dimension random vector ($24$ time periods for each wind farm). 
Further configuration of the IEGS (i.e., the power and gas network data, the wind power data, and the load profile) can be found on: \url{https://github.com/TxiaoWang/DDSPCE-based-Stochastic-ED.git}. \color{black} \color{black}

To test the performance of the proposed DDSPCE-based surrogate \color{black} model, we compare the probabilistic characteristics (e.g., mean $\mu$, standard deviation $\sigma$, the PDF and the CDF) of the minimum cost $Q$ estimated from the DDSPCE-based surrogate \color{black}model with  those by the SPCE method \cite{Sheng2018} (e.g., with inferred PDF \cite{Torre2021}) \color{black} and those from the benchmark $10,000$-sample MC simulations. 
Note that the SPCE-based model in this simulation is built based on the PDF inferred from the available data while the proposed method is built directly from data\color{black}.  It can be seen in 
Table \ref{tab:IEEE118_cost} and Fig. \ref{fig:Obj_comp} 
that  both the proposed DDSPCE method and the SPCE method can provide good estimations for the probabilistic characteristics of the minimum cost $Q$, though the proposed method possesses a better accuracy in the estimations.  
\color{black}

Particularly, the DDSPCE-based surrogate \color{black} model (\ref{eq:PCEinfty}) is constructed using only $1,100$ samples  (e.g., number of training sample $\mathcal{N} = 1100$)\color{black}. The time consumption of the DDSPCE method is only about $\frac{1}{9}$ of the time consumed by the MC simulations. However, if only mean and variance are needed while the detailed PDF is not of interest, only $300$ samples  ($\mathcal{N}=2.5M$) \color{black} are required 
to achieve accurate estimations (see Table \ref{tab:IEEE118_cost_N300})\color{black})
, which takes only about $\frac{1}{33}$ of the time required by the MC simulations.  Besides, the SPCE method takes $8.83s$ more than the proposed method due to the additional PDF inferring procedure. 
\color{black} 
\vspace{-0.26in}
\begin{table}[!ht]
\setlength{\abovecaptionskip}{-0.35cm}
\setlength{\belowcaptionskip}{-1.4cm}
\renewcommand{\arraystretch}{1.0}
\small
\caption{Comparison of the estimated statistics of 
$Q$ by the MC simulations, the DDSPCE and the SPCE methods with $\mathcal{N}=1100$. \color{black}}
\label{tab:IEEE118_cost}
\centering
\begin{tabular}{c|c|c|c}
\hline
Index                                                 & MC                 & DDSPCE                    &   SPCE \color{black}  \\ \hline
$\mu$                                                 & $7.3276 \times 10^6$ & $7.3276 \times 10^6$    &   $7.3277 \times 10^6$  \color{black}        \\ \hline
$\sigma$                                                  & $4.7069\times 10^4$  & $4.7159\times 10^4$ &     $4.8600\times 10^4$  \color{black}       \\ \hline
$\frac{\Delta \mu}{\mu_{\mathrm{MC}}}\%$          & $--$                 & $\bm{2.0739\times 10^{-5}}$      &     $1.8503\times 10^{-3}$  \color{black}         \\ \hline
$\frac{\Delta \sigma}{\sigma_{\mathrm{MC}}}\%$ & $--$                 & $\bm{1.9218\times 10^{-1}}$         &     $3.2529$ \color{black}         \\ \hline
\end{tabular}
\vspace{-6pt}
\end{table}
\normalsize
\begin{figure}[!ht]
\vspace{-12pt}
\setlength{\abovecaptionskip}{-0.3cm}
\setlength{\belowcaptionskip}{-0.6cm}
\centering
\includegraphics[width=0.26\textwidth]{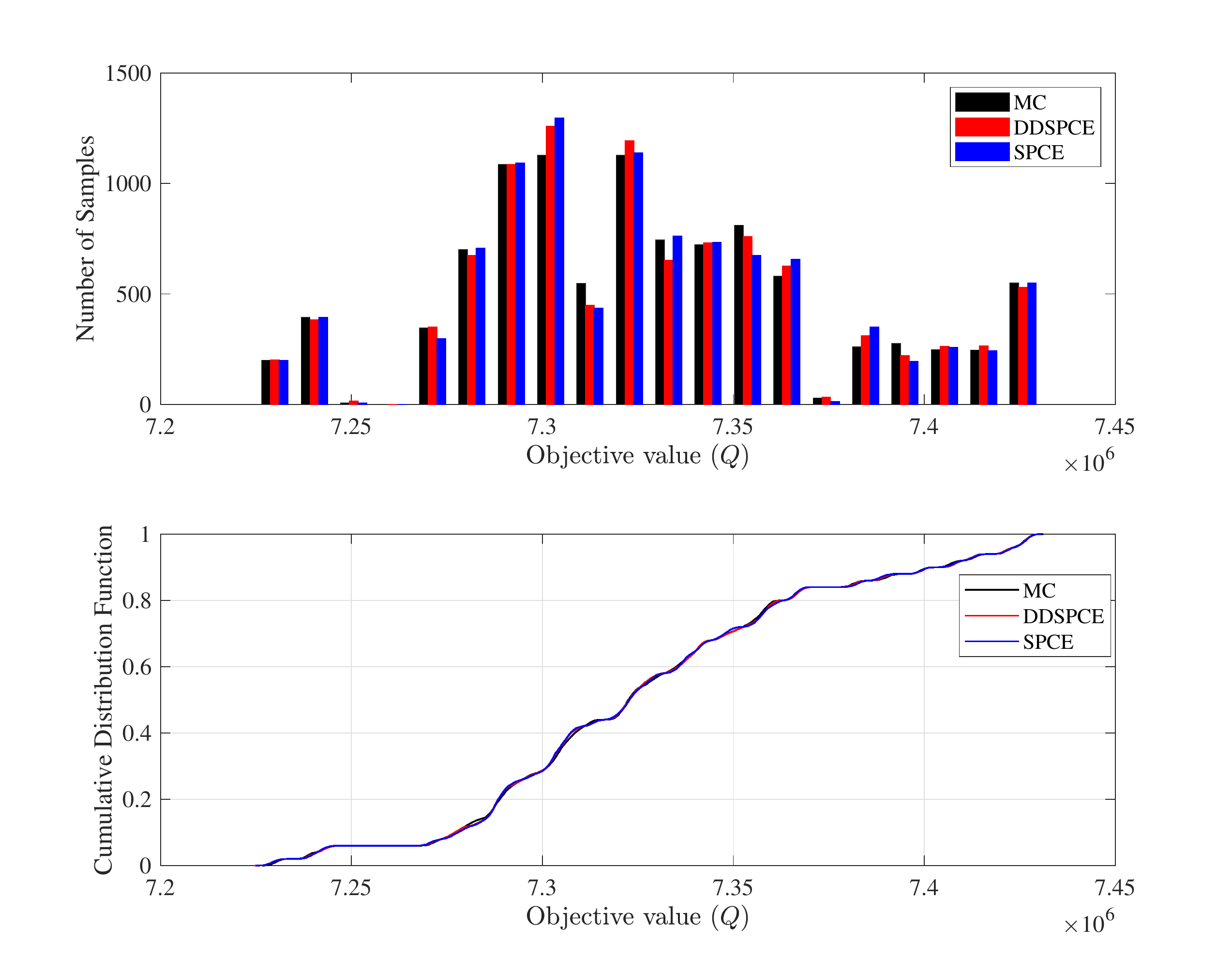}
\caption{Comparison of the PDF and CDF of the minimum cost $Q$ by the MC simulations,  the DDSPCE, and the SPCE methods with $\mathcal{N}=1100$.\color{black}}
\label{fig:Obj_comp}
\end{figure}
\vspace{1pt}

\noindent\textbf{Remark: } From our simulation experience, 
$\mathcal{N}\approx$ $2.5M$ is typically sufficient
for accurate mean and variance. 
While for an accurate PDF estimation for a \textit{unimodal} system response, 
$\mathcal{N}\approx 5M$ is typically required. 
For an accurate PDF estimation for a \textit{multimodal} system response, more samples (e.g., $\mathcal{N}\approx 9M$ in this letter) may be required. 
\color{black}
\vspace{-14pt}
\begin{table}[!ht]
\setlength{\abovecaptionskip}{-3pt}
\renewcommand{\arraystretch}{1.0}
\small
\caption{Comparison of the estimated statistics of 
$Q$ by the MC simulations, the DDSPCE and the SPCE methods with $\mathcal{N}=300$. \color{black}}
\label{tab:IEEE118_cost_N300}
\centering
\begin{tabular}{c|c|c|c}
\hline
Index                                                 &  MC                 &  DDSPCE                    &   SPCE \color{black}  \\ \hline
 $\mu$                                                 &  $7.3276 \times 10^6$ &  $7.3276 \times 10^6$    &   $7.3277 \times 10^6$  \color{black}        \\ \hline
 $\sigma$                                                  &  $4.7069\times 10^4$  &  $4.7228\times 10^4$ &     $5.0379\times 10^4$  \color{black}       \\ \hline
 $\frac{\Delta \mu}{\mu_{\mathrm{MC}}}\%$          &  $--$                 &  \bm{$4.8945\times 10^{-4}$}      &     $1.7092\times 10^{-3}$  \color{black}         \\ \hline
 $\frac{\Delta \sigma}{\sigma_{\mathrm{MC}}}\%$ &  $--$                 &  $ \bm{3.3939\times 10^{-1}}$         &     $7.0318$ \color{black}         \\ \hline
\end{tabular}
\vspace{-6pt}
\end{table}
\arrayrulecolor{black}
\vspace{-1pt}
\section{Conclusions}
\vspace{-3pt}
In this letter, we propose a data-driven sparse PCE (DDSPCE)-based \color{black}surrogate model 
for the SED problem. The probability characteristics of the SED solution (i.e., minimum production cost) can be approximated by the DDSPCE-based surrogate \color{black} model without pre-assumed probability distribution of random inputs. Simulation results on  an  IEGS 
system (the IEEE 118-bus system integrated with a 20-node \color{black} gas system) 
verify that the proposed DDSPCE method can provide accurate estimations for the mean, variance, PDF and CDF of the SED solution accurately and efficiently, even when the probability distribution of the SED solution is multimodal.

Particularly, compared to the method in   \cite{Wang2021}, orthogonal matching pursuit (OMP) is applied in this letter to find the coefficients of the DDSPCE method, which may achieve a faster convergence than the method in  \cite{Wang2021}. Besides, unlike in  \cite{Wang2021} where both the random inputs and the model response (Total Transfer Capability) are unimodal, the random inputs from real-world data and the model response (the minimum cost $Q$) of this letter are both multimodal, showing the accuracy and efficiency of the  DDSCPE method in handling more generalized situations.

\appendix \label{appendix}
For the IESG system, 
the multi-period SED problem under fixed UC decisions can be represented by:
\small{
\begin{equation}
\setlength{\abovedisplayskip}{2pt}
\setlength{\belowdisplayskip}{-1pt}
\label{eq:SED_Obj_gas}
   Q(\bm{P_g},\bm{P}_{\bm{w}})= \min_{\bm{P_g}} \sum_{t\in T} \left(\sum_{g\in G} C_{g}(P_{g}^{t}) + \sum_{s\in S} C_{s}(g_{s}^{t})\right)
\end{equation}   
}\normalsize
\small{
\begin{subequations}
\vspace{-0.12in}
\setlength{\belowdisplayskip}{2pt}
\label{eq:constraints_gas}
\begin{align}
&\mathrm{s.t.}  \notag \\
& \mathrm{Constraints} \ \mathrm{ (\ref{eq:Power_bal})-(\ref{eq:Ramp}) }\notag \\
& G_{s}^{\mathrm{min}}\leq g_{s}^{t}\leq G_{s}^{\mathrm{max}} \quad \forall s\in S,  t\in T \label{eq:gas_cons}\\
& G_{a}^{\mathrm{min}}\leq \pi_{a}^{t}\leq G_{a}^{\mathrm{max}} \quad \forall a\in G_a,  t\in T \label{eq:gas_consn} \\
& g_{b}^{t} = W_{b}\sqrt{((\pi_{e(b)}^{t})^{2}-(\pi_{a(b)}^{t})^2)} \quad \forall b\in G_b, t \in T \label{eq:eq:gas_consb} \\
& \pi_{e(c)}^{t}\leq \alpha_{c}\pi_{a(c)}^{t} \quad \forall c \in G_c, t\in T \label{eq:gas_conse} \\
& 0\leq g_{b}^{t}\leq G_{b} \quad \forall b \in G_b, t\in T \label{eq:gas_cons_Gb} \\
& 0\leq g_{c}^{t}\leq G_{c} \quad \forall c \in G_c, t\in T \label{eq:gas_cons_Gc} \\
& \sum_{s\in G_s} g_{s(a)}^{t}+\sum_{b_1 \in G_b} g_{b_1(a)}^{t}- \sum_{b_2 \in G_b} g_{b_2(a)}^{t} +\sum_{c_1\in G_c}g_{c_1(a)}^t \notag \\
&-\sum_{c_2\in G_c}g_{c_2(a)}^t = \sum_{d \in G_d} G_{d(a)}^t+\sum_{g\in G} \Theta_{g}P_{g(a)}^{t} \quad  t\in T \label{eq:gas_cons_Geq} 
\end{align}
\end{subequations}
}
\normalsize
where 
$C_s(\centerdot)$ is the cost of gas well. 
Gas network constraints are shown in (\ref{eq:constraints_gas}), 
where (\ref{eq:gas_cons}) denotes the output capacity limits of a gas well;  constraint (\ref{eq:gas_consn}) denotes the nodal pressure range; 
constraint (\ref{eq:eq:gas_consb}) describes the gas flow $g_b^t$ in gas passive pipeline $b$; 
constraint (\ref{eq:gas_conse}) is the simplified gas compressor model; 
$a(c)$; 
constraints (\ref{eq:gas_cons_Gb})-(\ref{eq:gas_cons_Gc}) denote the gas flow transmission capacity limits in a gas passive pipeline and gas compressor, respectively; 
constraint (\ref{eq:gas_cons_Geq}) denotes the gas nodal balance. The detailed notations can be found in \cite{Liu2021}. 
\color{black}

\end{document}